# From the No-Signaling Theorem to Veiled Nonlocality

Subhash Kak[1]

**ABSTRACT**
According to the no-signaling theorem, the nonlocal collapse of the wavefunction of an entangled particle by the measurement on its twin particle at a remote location cannot be used to send useful information. Given that experiments on nonlocal correlations continue to have loopholes, we propose a stronger principle that the nonlocality of quantum mechanics itself is veiled. In practical terms, decoherence and noise compels us to view the wavefunction as representing knowledge of potential outcomes rather than the reality. Experimental evidence in favor of naked nonlocality would support the view of the wavefunction as an objective description of physical reality.

## INTRODUCTION

The nonlocality of quantum mechanics, as expressed, for example, by the property of entanglement between remote particles, is consistent with the no-signaling theorem: "through quantum nonlocality there will be no way to exert instant control over what happens at far away places, nor to transmit signals to such places." [1, page 158]: This theorem mitigates the observational effects of nonlocality and "in the context of large scale level in which physical investigations are carried out, nonlocal effects are not significant… [and it allows] for separability of systems." [1, page 157] Nevertheless, experiments to investigate nonlocality have been designed, but imperfections in the experiments leave various loopholes permitting explanations based on local realistic theories [2]-[4]. Is it possible that the loopholes will never be closed and experimental verification of nonlocality that excludes local realistic explanations will not be found? If that is the case then there should be a principle stronger than that of no-signaling and it may be called the principle of *veiled nonlocality*.

Einstein, Podolsky, and Rosen (EPR) argued from the philosophical position of realism [5] that quantum mechanics could not be considered a complete theory since for an entangled pair of particles that are far apart from each other, the measurement on one causes the second to change its wavefunction. This influence projected nonlocally across what could be a vast distance is against our commonsense expectation of a physical process. Although EPR did not use the phrase "nonlocality" in their paper, it is clear that this feature of quantum mechanics was behind their assertion that it could not be a complete theory. As a realist one cannot accept that a measurement at one location, which in principle is arbitrarily far from the second location, can influence an object at this second location.

Bell considered the question if the property being measured was fixed at the time the pair of entangled particles was produced or whether the random collapse took place at the time of measurement after the particles had separated. He showed that if measurements were made independently in three different directions by Alice and Bob, the constraints on probability for the quantum case were different than for the classical case [6] (see also [7] for an overview and a slightly different formulation). Bell's theorem is taken to mean that quantum theory cannot be mimicked by introducing a set of objective local "hidden" variables. It follows that any classical theory advanced in place of quantum mechanics will be nonlocal.

---

[1] Oklahoma State University, Stillwater, OK 74078, USA.

Those who approach the subject from the position of realism offer the possibility that the collapse of the state function of two remotely situated entangled objects will leave some trace in terms of local process explanations that will call for a theory that goes beyond current quantum theory [8],[9].

In the two-slit experiment, we visualize the wavefunction of the particles going through both the slits and what collapses when the wave function *strikes* the screen is the probability amplitude. It is assumed that nothing physical is traveling faster than the speed of light or going through the slits and no messages or signals can be sent using this collapse of probability. When two particles are entangled as in $|\varphi\rangle = \frac{1}{\sqrt{2}}(|00\rangle + |11\rangle)$ and they are spatially separated, the particles are each in a mixed state. The measurement of one in a particular orientation leads to an identical result in the remote location if it is carried out in the same orientation. If the measurements are made in different orientations, the quantum system gives results that are different from the classical case. In principle, this difference can be used to determine if the process under study is classical or quantum, but the derivation is based on certain assumptions that may not be easy to satisfy in practical implementations.

The measurement problem can be viewed from a variety of perspectives:

1. The "reductionist" language used to explain the observations is not appropriate which is why we have the paradox that each particle is seen to go through both slits. According to the Copenhagen Interpretation, quantum theory is concerns our knowledge of reality and not the structure of reality and one can only speak of a probabilistic final result that is a function of the initial conditions.

2. The universe is interconnected and random processes are not quite random and local. Reality is characterized by nonlocality and undivided wholeness and the collapse is only apparent and it is triggered by the interaction with the environment. According to Bohm's and Hiley's ontological interpretation of reality [1, page 179]: "The overall quantum world measures and observes *itself*. For the classical 'sub-world' that contains the apparatus is inseparably contained within the subtle quantum world, especially through those nonlocal interactions that bring about the classical behavior. In no sense is the 'observing instrument' really separate from what is observed."

3. Matter at the microscopic level is not "physical" and the only reality is mathematical, in terms of probability amplitudes [10]. In such a realistic interpretation, nonlocality should have measurable consequences.

In principle, macroscopic systems may also be quantum and the quantum/classical divide of the Copenhagen Interpretation is somewhat vague. It seems reasonable to take measurement process to be decoherence caused by the environment [11] and, therefore, there is no need to seek the agency of consciousness in the transition from the quantum to the classical world, as was suggested by scientists such as Wigner and Stapp [12],[13]. On the other hand our conception of the universe and its laws and our analysis is a result of the capacities of our cognitive systems [14],[15]. Although it is possible that the classical agents of our mind are based on quantum collectives [16], we cannot negate the fact that the world of perceptions is at the basis of our knowledge and this world has a classical form.

Those who are dissatisfied with the Copenhagen Interpretation suggest new way of looking at things such as the Many Worlds Interpretation. Other less popular interpretations are consistent histories and modified dynamics. Another possible explanation is in some appropriate version of hidden-variable theory. Yet another view is to assume that there is a deeper theory that agrees with the quantum theory in the microscopic world and the classical theory in the macroscopic world.

By the no-signaling theorem, the "instantaneous" collapse of the wavefunction of an entangled particle by the measurement on its twin particle at a remote location cannot be used to send useful information, but experiments on nonlocal correlations continue to have loopholes and therefore they cannot be considered as definitive tests of nonlocality. The two principal loopholes are those of *detection* and *locality*.

The detection loophole addresses the fact that although derivation of the Bell Inequality assumes binary outcomes, say 1 and -1, in reality a third outcome of "no-click" is associated with the observations. Furthermore, as practical detectors are not perfectly efficient, the "no click" data cannot be left out as being anomalous under a fair sampling assumption. For the low detection efficiency case, the experimental results can be explained by a local realistic theory [4], [17]-[19].

The locality loophole addresses the possibility that a local realistic theory might rely on some type of slower-than-light signal sent from one entangled particle to its partner. Furthermore, the measurement choice on one side should not be correlated with that at the other. This has also been variously called the "freedom-of-choice" and the "measurement-independence" loophole. According to Brunner et al [4]:
> [C]ontrarily to the detection loophole, the locality loophole can never be "completely" closed. Strictly speaking, it requires spacelike separation between the event determining the choice of measurement setting on one side and the event corresponding to the output of the measurement device on the other side. The problem is that this requires to know precisely the time at which these two events happen. But if we use some random process that outputs a random value at time, say t = 0, how do we know that this value, even if it is independent from the hidden variable and from the mechanism determining the measurement setting on the other side, was precisely generated by the random process at this time and not at some earlier time t = −δ? Similarly, how do we know precisely when a measurement is completed without making some assumptions on the collapse of the wavefunction?

Will future experiments bridge these loopholes and demonstrate nonlocality that cannot be explained by a local realistic theory? Nearly a decade ago, Santos summarized the situation thus [19]: "the forty-years failure to show a strict (loophole-free) violation of a Bell inequality may be 'explained' by another fundamental principle of physics: *nature respects local realism.*" The assessment of Santos still remains true but rather than taking this to indicate a local realistic basis to quantum theory, we argue that it implies that nonlocality is veiled which is consistent with the Copenhagen Interpretation. Our mind operates by the classical picture and it uses artifacts wherever necessary [20],[21] and the principle makes the classical picture of reality remain consistent. If the principle were true, local explanations will be possible in experiments seeking nonlocality and distant correlations will be coincidences.

## VEILED NONLOCALITY AND THE NATURE OF THE WAVEFUNCTION

Veiled nonlocality facilitates choice between the two main interpretations of the wavefunction: (i) it represents objective reality in a mathematical form, and (ii) although sometimes the wavefunction describes the system fully, in most cases it only encodes information about the potential outcomes of the experiment together with their probabilities. If the wavefunction has ontological reality, then it is conceivable that an experiment will show the nonlocality, but if the wavefunction only encodes information about outcomes then it is unlikely that nonlocality will be revealed.

To review the measurement process, note that observable quantities are represented by Hermitian operators, and their possible values are the eigenvalues of these operators. Corresponding to the eigenvector $u_n$ of the pure state $\varphi$ the probability of detecting eigenvalue $\lambda_n$ is $|\langle u_n | \varphi \rangle|^2$. If the source output be represented by the positive operator $\rho$, then a complete set of positive operators $E_\mu$ (for various measurements $\mu$), which includes the no detection possibility, defines the positive operator-valued measure (POVM), and the probability $p(\mu) = \text{tr}(E_\mu \rho)$. For an initial pure state $\sum_i a_i |i\rangle$ and the apparatus (environment) $|A\rangle$, the joint state is initially $\sum_i a_i |i\rangle \otimes |A\rangle$ which evolves into the composite state $\sum_{i,w} a_i |i\rangle U_{i,w} |w\rangle$ where $|w\rangle$ is the basis for the composite state and $U_{i,w}$ is a suitable unitary operator.

The act of observation reduces the wavefunction to one of its component states and it is the outcome associated with the application of the measurement operator on the state [22],[23]. From the philosophical perspective the reduction of the wavefunction in a random fashion is a feature very different from that of its evolution although this jump may be viewed as not taking place in the physical world but rather in our knowledge of the system. Physically, the reduction can be viewed as decoherence precipitated by the environment [11],[24], or it can be avoided by speaking of complementarity. But the fact that single electrons also exhibit wavelike interference [25],[26] makes it right to ask the question as to how this complementarity plays out at the boundary between different types of behavior.

In one of the first interpretations of quantum mechanics, Bohr proposed that the measurements from the state function could be understood in a relational sense as in relativity theory. Later he and others argued for a positivist interpretation in which it is wrong to ask what the attributes of the object are prior to measurement and one can speak only of the observed values and this became the Copenhagen Interpretation (CI) [26]. Heisenberg insisted that "physics must confine itself to the description of the relationship between perceptions."

In the CI, there is a split between the classical world of the observers and the quantum microworld that is being observed. Bohr insisted [27]: "In the system to which the quantum mechanical formalism is applied, it is of course possible to include any intermediate auxiliary agency employed in the measuring process [but] some ultimate measuring instruments must always be described entirely on classical lines, and consequently kept outside the system subject to quantum mechanical treatment."

Although CI is still the most widely interpretation used, it has been challenged in recent years by the Many Worlds Interpretation (MWI) that started off as a way to view the collapse at measurement in terms of random process theory in which a specific time series is viewed as just one member of a large ensemble. For a random process any specific realization need not provide all the information about its probability characteristics, but across the ensemble each time instant defines a random variable. If the properties of the random variables do not change with time then the process is stationary. Furthermore, if the properties across time are the same as properties across the ensemble, then the process is ergodic. MWI took the Born probability distribution to be literally true and, therefore, postulated other worlds (constituting the ensemble) that, together with our own, validate the observations.

Although initial descriptions of MWI took the world itself to split into many copies at each measurement (to satisfy ensemble characteristics), more recent versions of MWI take a somewhat different tack [28]. The wavefunction of the universe is now taken as the starting point in the consideration of reality and given it, there can be no collapse as is assumed in CI. The reduction of the wavefunction is replaced by decoherence induced by the environment which destroys superpositions in the macro scale.

If CI is the view of the universe with the observer in the privileged position, then MWI is the outside realistic view. In CI the wavefunction represents the knowledge of the experimenter, whereas in MWI the wavefunction is the complete reality. If CI is the subjective view, MWI is the objective view.

The inside-out and the outside-in are like the complementary wave and particle viewpoints already considered in CI. The outside-in view of MWI might present a consistent picture but it means that observers are zombies and in this it is similar to a conception of reality as nothing more than a collection of *things*. In such a picture, there is no room for minds with agency and the whole universe operates as a giant computer. In the MWI view the choices are determined completely by the environment.

The inside-out view of CI admits the possibility that somehow "free will" plays a role in the choice that emerges during a measurement (of course this narrative is relevant only in a limited set of possibilities). Since the cut between the classical and the quantum may be made at different points, this does not rule out the explanation of the outcomes of an experiment as decoherence by the environment.

The problem of measurement is linked with the idea of "free will". If the brain is viewed as a neural machine, its response to a stimulus is determined by its current internal state and therefore it cannot act acausally and exercise free will. "Consciousness" (which the mainstream view takes as an emergent phenomenon) then provides a false sense of agency by assigning its intervention an earlier time than is correct, as indicated by the experiments of Libet and others? [29]-[32]

The general position in the physics community on the question of the wavefunction was described by Peres and Terno in the following words [33]:

> Many physicists, perhaps a majority, have an intuitive realistic worldview and consider a quantum state as a physical entity. Its value may not be known, but in principle the quantum state of a physical system would be well defined. However, there is no experimental evidence whatsoever to support this naive belief. On the contrary, if this view is taken seriously, it may lead to bizarre consequences, called "quantum paradoxes." These so-called paradoxes originate solely from an incorrect interpretation of quantum theory. The latter is thoroughly pragmatic and, when correctly used, never yields two contradictory answers to a well posed question. It is only the misuse of quantum concepts, guided by a pseudorealistic philosophy, that leads to paradoxical results.

Going beyond physics, if a metaphysical element is injected into the discussion, "awareness" as "witness" is universal consciousness which by observation by the quantum Zeno effect guides unfoldment of the universe [34],[35]. This is similar to the CI perspective that provides the possibility, at least to the philosopher of mind who is looking for space where human agency might be located, for an acausal influence on a process. Such an argument is of no interest to the empiricist for as described by Moritz Schlick in his essay on the difference between positivism and realism [36]: "The denial of the existence of a transcendent external world would be as much a metaphysical statement as its affirmation. Hence the consistent empiricist does not deny the transcendent world, but shows that both its denial and affirmation are meaningless."

In contrast to CI, MWI is the machine view of the universe and it has no place for human agency. From subjective experience, the conscious individual is convinced that such an agency exists and other sentient beings also possess it. The various differences in CI and MWI are summarized in Table 1.

Table 1: Differences between the Copenhagen Interpretation (CI) and the Many Worlds Interpretation (MWI)

|  | CI | MWI |
|---|---|---|
| Wavefunction | State of knowledge of the system | Mathematical representation of reality |
| Measurement | Change in the knowledge of the system based on observations made | Decoherence of the system induced by the environment |
| Observers | Create representations of the universe | Observers merely record results; observers are like zombies |
| Reality | Constructed out of measurements | Reality exists outside of agents |
| Universe | One | Multiverse |
| Consciousness | Unitary | Many minds |

Many points of difference between CI and MWI that are of a philosophical nature do not have any implications as far as physical experiments are concerned. In a universe only of "things" which excludes "being", the problem of "free will" is unsolvable and in a theoretical framework based on causality, "free will" is paradoxical. The last row of Table 1 will be considered to be outside of the realm of physics by some and others would contest the characterization of consciousness in CI to be unitary although it was so assumed by Schrödinger [37, pages 170-173].

When the question of free will is raised in the philosophical foundations of physics, it is generally done assuming that observers are present in the universe, without explaining how such sentient agents arise. From the perspective of sentient beings, objects are not primary; what is primary are mental feelings and perceptions. Newer theories of physics consider information as primary, but there can be no information unless there is consciousness that can recognize this information. This consciousness (or being) may exist in a plane different from regular spacetime.

Accepting mental states as primary building-blocks of reality leads to questions such as: Are the worlds of mind (consciousness) and matter separate? If they are, how can the one influence the other? In the (Vedantic) model of consciousness as interpenetrating the material world, consciousness guides the evolution of the material world indirectly [37], but that is irrelevant as far as physics is concerned.

Consider now a known wavefunction that has been prepared in the laboratory. It has extension, that is, it is described across space and time. Since this extension collapses into a local attribute, what is the process by which this collapse takes place and might this be apprehended by our instruments? For entangled pair of particles that are far apart, we will have to postulate processes that travel faster than the speed of light, contradicting relativity. If instantaneous information transfer cannot take place, is it still possible to see correlations after the process has occurred that can only be explained by means of nonlocal processes? *The principle of veiled nonlocality* (PVN) ensures that we will be able to advance local explanations for the process or ascribe the results to coincidences. The veiling of the nonlocality occurs due to decoherence and noise.

Why can't PVN be negated when the wavefunction of the objects is known? Even for the case where the wavefunction is known, there are imponderables. Consider the state associated with two entangled electrons: $|\varphi\rangle_{AA} = \frac{1}{\sqrt{1+r^2}}(r|\uparrow\downarrow\rangle - |\downarrow\uparrow\rangle)$, where $|\uparrow\rangle$ represents the spin-up state and $|\downarrow\rangle$ represents the spin-down state. If a physical process has been used to generate such an entangled state, the value of $r$ would be unknown; indeed there may also be an unknown phase value associated with the wavefunction. On the other hand, it is not clear whether quantum gates can be implemented to generate this state with complete certainty [38]. In reality, therefore, the assumed state function $|\varphi\rangle_{AA} = \frac{1}{\sqrt{2}}(|\uparrow\downarrow\rangle - |\downarrow\uparrow\rangle)$ represents knowledge about the system in a collective sense for an ensemble of such particles would average out the results over a wide range of $r$ values. Thus, for example, when it is assumed that the product of a circuit provides the entangled pair of photons or in the method of generating maximally-polarization-entangled states by type-II spontaneous parametric down-conversion in Bragg reflection waveguides [39], the expression $|\varphi\rangle = \frac{1}{\sqrt{2}}(|00\rangle + |11\rangle)$ could be shorthand for

$$|\varphi\rangle = \frac{1}{\sqrt{1+r^2}}(r|00\rangle + |11\rangle) \tag{1}$$

In principle, one can prepare a quantum state precisely, but in reality there exist problems related to initialization and lack of knowledge of the environment and decoherence [40] that affects the state of the system in uncertain and uncontrollable ways. This means that uncertainty relations will continue to be central in the analysis of the evolution of the system.

## DETECTION LOOPHOLE AND THE CLASSICAL/QUANTUM DIVIDE

We consider the question of the detection loophole. Let for three events A, B, and C, the probability $P_{sc} = P_{same}(A,B) + P_{same}(A,C) + P_{same}(B,C)$ and $P_{nc} = P_{notsame}(A,B) + P_{notsame}(A,C) + P_{notsame}(B,C)$ where $P_{notsame}(A,B)$ means that the outputs A and B are different, etc, and the "c" in the subscript for $P_{sc}$ and $P_{nc}$ refers to "classical" as in classical probability. For the quantum case we speak of $P_{sq}$ and $P_{nq}$ as the corresponding probabilities. In an earlier paper [7], we showed that

$$P_{sc} = P_{same}(A,B) + P_{same}(A,C) + P_{same}(B,C) \geq 1 \tag{2}$$

for a classical system where each of the events A, B, and C is binary, and

$$P_{sq} = P_{same}(A,B) + P_{same}(A,C) + P_{same}(B,C) = \frac{1}{2} + \frac{5r^2 - 6r + 5}{8(1+r^2)} \tag{3}$$

for a quantum system with two outputs under certain conditions where r is the measure of non-maximal entanglement. It was further found that in the special case where $P_{same}(A,B,C) = 0$, $\frac{P_{sc}}{P_{nc}} = \frac{1}{2}$. If $P_{same}(A,B,C)$ is not zero and the probabilities of all the outcomes associated with the three events are

the same, then

$$\frac{P_{sc}}{P_{nc}} = 1 \tag{4}$$

For the quantum case, when $r = 1$, we have (see equations (30) and (31) of [7] for $r = 1$)

$$\frac{P_{sq}}{P_{nq}} = \frac{1}{3} \tag{5}$$

Whether a process is classical or quantum can be checked by the two equations (4) and (5) that govern corresponding probabilities. But since in an actual detection system there are situations when there are no clicks, we should relax the conditions under which (2) and (4) were derived.

But before we do that, let us consider the case where the events are associated with m different outcomes, 0, 1, 2, …, m-1. For the case of m=3, the following table sums up the situation:

Table 1. Probability of outcomes for events for m=3

| Index | A | B | C | Probability |
|---|---|---|---|---|
| 0 | 0 | 0 | 0 | p(0) |
| 1 | 0 | 0 | 1 | p(1) |
| 2 | 0 | 0 | 2 | p(2) |
| 3 | 0 | 1 | 0 | p(3) |
| 4 | 0 | 1 | 1 | p(4) |
| 5 | 0 | 1 | 2 | p(5) |
| 6 | 0 | 2 | 0 | p(6) |
| 7 | 0 | 2 | 1 | p(7) |
| 8 | 0 | 2 | 2 | p(8) |
| 9 | 1 | 0 | 0 | p(9) |
|  | . | . | . | . |
| 26 | 2 | 2 | 2 | p(26) |

It is quite clear that $P_{same}(A,B)$ will consist of 9 terms out of 27. Assume that the probability of each of the terms is identical and equal to $p = 1/27$ then $P_{same}(A,B) = 1/3$. Therefore, $P_{same}(A,B) + P_{same}(A,C) + P_{same}(B,C) = 1$, and $P_{notsame}(A,B) + P_{notsame}(A,C) + P_{notsame}(B,C) = 2$. We can write down the following general result:

**Theorem**. For three events A, B, C, with each event associated with one outcome out of 0, 1, 2, … m-1 where the probability of each outcome is $p = 1/m^3$, we have:

$$\frac{P_{sc}}{P_{nc}} = \frac{1}{m-1} \tag{6}$$

*Proof.* It is clear that $P_{sc} = P_{same}(A,B) + P_{same}(A,C) + P_{same}(B,C) = 3m^2 \times p = 3/m$ and $P_{nc} = P_{notsame}(A,B) + P_{notsame}(A,C) + P_{notsame}(B,C) = 3(m-1)/m$, from which the result (6) follows.

It is striking that the result (6) for classical systems for $m = 4$ is identical to the result (5) for quantum systems for binary outputs. One way to interpret this is to see the quantum case as having twice the dimensions of the classical case by virtue of phase.

Now consider that each event is associated with three outcomes: 0, 1, and 2 (namely "no-click"). We will consider two special cases:

*Case 1.* Assume that while we associate each outcome with the same probability, we will not count events where there is a "no click". Given that, if we refer to Table 1,

$$P_{same}(A,B) = p(0)+p(1)+p(2)+p(12)+p(13)+p(14) \qquad (7)$$

We see that $P_{notsame}(A,B)$ also has six terms, and if it is assumed that each of these terms is identical, then $P_{same}(A,B)/P_{notsame}(A,B) = 1$. Since the situation is symmetric for all of the three pairs of events, we can write that $\frac{P_{sc}}{P_{nc}} = 1$.

*Case 2.* We associate each outcome with the same probability, but we only count events where there is at least one valid signal.

$P_{same}(A,B)$ remains the same as in Case 1 above. But $P_{notsame}(A,B) = p(3)+ p(4)+ p(5)+ p(6)+ p(7)+ p(8)+ p(9)+ p(10)+p(11)+ p(15)+ p(16)+ p(17)+ p(18)+p(19)+ p(20)+ p(21)+ p(22)+ p(23)$. Due to symmetry, the cases for the event pairs (A,C) and (B,C) will be similar. If the event probabilities are all identical, then since the number of terms of "same" are only one-third as many as in the "not-same" events,

$$\frac{P_{sc}}{P_{nc}} = \frac{1}{3} \qquad (8)$$

which is identical to the quantum case.

## CONCLUSIONS

The paper argues that while collapse of the state function in quantum mechanics is best described as a nonlocal process, this nonlocality is veiled. Probability constraints that exclude local realism explanations for quantum phenomena have loopholes in practical implementations (e.g. [17],[18],[41]).

If the case is made that entanglement or nonlocality is at the basis of the superiority of quantum computing algorithms for some problems, then the effective veiling of this nonlocality will preclude its exploitation in any practical implementation scheme. In this we agree with the assessment of Santos [18]: "If nature respects local realism it is far from obvious that applications of quantum physics relying on "purely quantum phenomena", like entanglement, could produce anything not achievable with devices working according the laws of classical physics. This might be, for instance, the case of quantum computation. If there is a general principle preventing the violation of a Bell inequality, then it is probable that the same principle might prevent the expected functioning of (large scale) quantum computers."

The veiling of nonlocality is due to decoherence and noise and it implies a stronger assertion than the no-signaling theorem. In practical terms, it compels us to view the wavefunction as representing our lack of knowledge rather than the reality. Any evidence in favor of naked nonlocality would support the view of the wavefunction as a physical reality.

Nonlocality lies outside the framework of classical science and, therefore, its actual occurrence is problematic in making sense of the world. We argue that veiled nonlocality, which appears to be consistent with the Copenhagen Interpretation, saves us from this difficulty.

*Acknowledgements.* This research was supported in part by the National Science Foundation grant CNS-1117068.